\documentclass[seceq]{ptptex}

\usepackage{graphicx}
\newcommand{\Z}{\mathbb{Z}}
\newcommand{\n}{\nonumber \\}
\newcommand{\tr}{\mbox{tr}}

\title{
On relationships among 
         Chern-Simons theory, \\
         BF theory and matrix model}

\author{
Takaaki
\textsc{Ishii}$^{1,}$\footnote{ishii@het.phys.sci.osaka-u.ac.jp},
Goro
\textsc{Ishiki}$^{1,}$\footnote{ishiki@het.phys.sci.osaka-u.ac.jp},
Kazutoshi
\textsc{Ohta}$^{2,}$\footnote{kohta@tuhep.phys.tohoku.ac.jp}, \\
Shinji
\textsc{Shimasaki}$^{1,}$\footnote{shinji@het.phys.sci.osaka-u.ac.jp} 
and
Asato
\textsc{Tsuchiya}$^{1,}$\footnote{tsuchiya@het.phys.sci.osaka-u.ac.jp}
}

\inst{
$^1$ Department of Physics, Graduate School of  
Science, Osaka University, \\
Toyonaka, Osaka 560-0043, Japan \\
$^2$  High Energy Theory Group, Department of Physics, Tohoku
University, \\  Sendai 980-8578, Japan
}

\abst{
Chern-Simons theory on a $U(1)$ bundle over a Riemann surface $\Sigma_g$ of
genus $g$ is
dimensionally reduced to BF theory with a mass term, which
is equivalent to the two-dimensional Yang-Mills  on $\Sigma_g$. We show that
the former is inversely obtained from the latter by the extended matrix 
T-duality developed in hep-th/0703021. For the case of 
$g=0$ (i.e. $S^2$), the $U(1)$ bundle represents the lens space 
$S^3/\Z_p$. We find that in this case both the Chern-Simons theory and
the BF theory with the mass term are realized in a matrix model.
We also construct Wilson loops in the matrix model that correspond to
those in the Chern-Simons theory on $S^3$.}

\begin{document}

\maketitle

\section{Introduction}
\setcounter{equation}{0}
\renewcommand{\thefootnote}{\arabic{footnote}} 
Some matrix models have been proposed as nonperturbative formulation of 
superstring or M-theory.\cite{BFSS,IKKT,DVV} \ Since the information of 
topology is relevant for compactification
in string theory, it should be included in these matrix models.
The topological field theories have been developed to efficiently
describe the topological aspects of field theories.
It is, therefore, worthwhile to investigate realization of 
the topological field theories in matrix models.

The large $N$ reduction \cite{EK} 
is the first example for realization of field theories
in matrix models. It states that a large $N$ planar gauge theory
is equivalent to the matrix model
that is its dimensional reduction to zero dimensions 
unless the $U(1)^D$ symmetry
is broken, where $D$ denotes the dimensionality of the original gauge theory.
However, the $U(1)^D$ symmetry is in general 
spontaneously broken in the case of $D>2$.
There are two improved versions of the large $N$ reduced model
that preserve the $U(1)^D$ symmetry.
One is the quenched reduced model.
\cite{Bhanot:1982sh,Parisi:1982gp,Gross:1982at,Das:1982ux} \ 
This shares the same idea with the T-duality for D-brane effective 
theories,\cite{Taylor:1996ik} \ which we call the matrix T-duality in this
paper. The other is the twisted reduced 
model,\cite{GonzalezArroyo:1982hz} \ 
which was later rediscovered in the context of 
the noncommutative field theories.  
The above developments are all concerning gauge theories on flat space-time.

In this paper, as typical examples of the topological field theories, we
consider three-dimensional
Chern-Simons (CS) theory on a $U(1)$ bundle over a Riemann surface
$\Sigma_g$ of genus $g$
and two-dimensional BF theory with a mass term on $\Sigma_g$ that 
is obtained by a dimensional reduction of the CS theory on the total space to 
the base space\footnote{This dimensional reduction was suggested in
\citen{Blau:2006gh}.}. The BF theory with the mass term is equivalent to the 
two-dimensional Yang-Mills on $\Sigma_g$.
Hence, for our purpose, we need to extend the prescription of 
the large $N$ reduction or the matrix T-duality
to that on curved space\footnote{An interesting
approach to description of curved spacetime by matrices was proposed
in \citen{Hanada:2005vr}.}. 
In \citen{ISTT,IIST}, the matrix T-duality was extended to 
the relationship between a Yang-Mills-type gauge theory 
on the total space of a $U(1)$ bundle and the Yang-Mills-Higgs theory
on its base space that is obtained by dimensionally reducing the original
theory with respect to the fiber direction.
Also, many works have been done on realization of 
the gauge theories on fuzzy sphere 
\cite{Madore:1991bw,Grosse:1992bm,Grosse:1995ar,CarowWatamura:1998jn} 
by matrix models
\cite{Iso:2001mg}
and the monopoles on
fuzzy sphere 
\cite{Grosse:1995jt,Baez:1998he,Landi,CarowWatamura:2004ct,Aoki:2003ye}.
It was explicitly
shown in \citen{ISTT} that if an appropriate continuum limit is taken,
the theory expanded around a multi-monopole
vacuum of the Yang-Mills-Higgs theory on $S^2$ is equivalent to the theory
expanded around a vacuum with concentric fuzzy spheres of the matrix model 
that is obtained by dimensionally reducing the original theory on $S^2$
.

We show that the CS theory on the $U(1)$ bundle
is inversely obtained from the BF theory with the mass term on the base space
by the extended matrix T-duality developed in \citen{IIST}. 
For the case of $g=0$ (i.e. $S^2$), the $U(1)$ bundle represents the lens space
$S^3/\Z_p$. We show that in this case both the CS theory and
the BF theory with the mass term are realized in the matrix model that is 
obtained by dimensionally reducing the both theories to zero dimensions.
We also construct Wilson loops in the matrix model that correspond to
those in the CS theory on $S^3$.
This is an extension of the work \citen{Ishibashi:1999hs} to curved space.

Our study of the relationship between the CS theory and 
the BF theory with the mass term is motivated by the following observations
on the relationship among the matrix models, topological string and
lower dimensional topological field theory.
The free energy of the hermitian (holomorphic) matrix model 
in the large $N$ limit can be regarded as a topological B-model 
amplitude on a suitable conical 
Calabi-Yau singularity 
\cite{Dijkgraaf:2002fc,Dijkgraaf:2002vw,Dijkgraaf:2002dh}.
The topological string amplitude obtained from the matrix model 
can also be applied to derivation of effective superpotential 
of four-dimensional $ {\cal N}=1$ supersymmetric gauge theory.
It is also argued that the ${\cal N}=1$ (quiver) matrix models ``deconstruct''
the CS and BF-type theory \cite{Dijkgraaf:2003xk}.
If we now demand a periodicity in the space of the matrix eigenvalues, 
the Vandermonde measure becomes the unitary one by taking account 
of mirror images of the eigenvalues (D-branes) in the covering space.
This unitarization is reminiscent of the (matrix) T-duality we have mentioned.
Indeed, it is shown that the matrix model with unitary measure is equivalent 
to the partition function of the CS theory, 
and then its free energy represents a topological A-model amplitude 
on a mirror Calabi-Yau geometry \cite{Aganagic:2002wv}. 
(See also for reviews \citen{Marino: 2004uf,Marino:2004eq}.)
Thus, these topological string amplitudes emerging from the matrix models are 
related with each other by T-duality and mirror symmetry in string theory.
On the other hand, two-dimensional Yang-Mills theory and its generalization, 
which is obtained by a potential deformation of the BF theory 
(generalized two-dimensional Yang-Mills \cite{Douglas:1994pq}), 
can be interpreted as a matrix model restricted on 
discrete eigenvalues \cite{Gross:1994ub,Matsuura:2005sg}. 
In particular, the two-dimensional Yang-Mills theory on $S^2$ 
possesses the same kind of the Vandermonde determinant, 
which is a square of the dimension of a unitary group representation, 
as the hermitian one matrix model.
Using the similar idea of the T-duality, the Vandermonde measure 
(the dimension of the unitary representation) is replaced with 
an unitary one, which is a $q$-analog of the dimension of 
the unitary representation. This model is called $q$-deformed 
two-dimensional Yang-Mills theory \cite{Aganagic:2004js}. 
(See also \citen{deHaro:2004id,deHaro:2004uz}.)
The $q$-deformed two-dimensional Yang-Mills theory includes 
non-perturbative aspects of the topological A model string. 
Actually, we can obtain the partition function of the CS theory 
on $S^3$ by extracting a ``ground state'' configuration from the 
$q$-deformed two-dimensional Yang-Mills theory.
These facts on the relationship between various matrix models and lower 
dimensional solvable gauge theories strongly suggests that there exists 
T-dual like relationship between two-dimensional Yang-Mills theory on 
$S^2$ and the CS theory on $S^3$.

This paper is organized as follows. In section 2, we show the relationship
between the CS theory on the $U(1)$ bundle over $\Sigma_g$
and the BF theory with the mass term
on $\Sigma_g$. In section 3, we show the relationship between BF theory
on $S^2$ and the matrix model. We also find how CS theory on $S^3/\Z_p$ is
realized in the matrix model. In section 4, we construct the Wilson loops
in the matrix model that correspond to those in CS theory on $S^3$.
Section 5 is devoted to summary and discussion. In appendix, some formulae
concerning the 
spherical harmonics on $S^3$, $S^2$ and fuzzy sphere are gathered.

\section{ CS theory vs BF theory with the mass term}
\setcounter{equation}{0}
We consider a $U(1)$ bundle ${\cal M}$ over a closed Riemann surface $\Sigma_g$
of genus $g$.
The base space $\Sigma_g$ has a covering ${\cal S}$,
and the total space ${\cal M}$ has a covering $\{\pi^{-1}(U)|U \in {\cal S}\}$. 
$\pi^{-1}(U)$ is diffeomorphic to $U\times S^1$ by
a local trivialization. Thus it is parameterized 
by $z^M=(x^{\mu},y)\;(M=1,2,3, \; \mu=1,2)$, 
where $x^{\mu}$ parameterize the local patch $U$
and $y$ parameterizes the $S^1$ with $0 \leq y \leq 2\pi R$.
If there is overlap between the local patches $U$ and $U'$, the relation
between $y$ and $y'$ is given by the transition function
$e^{-\frac{i}{R}v}$ as $y'=y-v(x)$.
We can assume that the total space is endowed with a metric which
can be expressed locally on $U$ as 
\begin{align}
ds^2=G_{MN}dz^Mdz^N
=g_{\mu\nu}(x)dx^{\mu}dx^{\nu}+(dy+b_{\mu}(x)dx^{\mu})^2,
\end{align}
where $g_{\mu\nu}$ is the metric of the base space. $b/R$ is 
the local connection that is viewed as the monopole on the base space
and transformed as 
$\frac{1}{R}b'
=\frac{1}{R}(b+dv)$. The curvature 2-form $db/R$ belongs to 
$H^2(\Sigma_g,\Z)$. From this fact we can assume that
\begin{align}
\frac{1}{2\pi R}b_{\mu\nu}=-\frac{p}{V}\sqrt{g}\epsilon_{\mu\nu},
\label{curvature 2-form}
\end{align}
where $b_{\mu\nu}=\partial_{\mu}b_{\nu}-\partial_{\nu}b_{\mu}$,
$p$ is an integer that represents the monopole degree of the $U(1)$ bundle and
$V$ is the volume of the base space. Note that (\ref{curvature 2-form}) is
consistent with the equality
\begin{align}
\frac{1}{2\pi R}\int_{\Sigma_g}db=-p.
\end{align}
In the following, all the expressions only
make sense locally on $U$ or $\pi^{-1}(U)$
unless any remark is made.

We next consider a gauge theory on the above 
total space and make a dimensional reduction
of the fiber direction to obtain a gauge theory on the base space.
The gauge field of the fiber direction, $A_y$, is identified with the Higgs
field on the base space. The horizontal-vertical decomposition given
by the connection 1-form $\omega=\frac{1}{R}(dy+b)$
tells us how to decompose the gauge fields on the total space into 
the gauge fields $a_{\mu}$ and the Higgs field $\phi$ 
on the base space \cite{IIST}\footnote{Eq.(4.8) in \citen{IIST} 
expressed in the local Lorentz frame is equivalent to (\ref{decomposition
of gauge fields}).}:
\begin{align}
A_{\mu}&=a_{\mu}+b_{\mu}\phi,  \nonumber\\
A_y&=\phi,
\label{decomposition of gauge fields}
\end{align}
where both sides of these equations are assumed to be independent of $y$.

We start with the $U(M)$ CS theory on the total space:
\begin{align}
S_{CS}&=\frac{k}{4\pi}\int_{\cal M} d^3z \:\epsilon^{MNP}
\mbox{Tr}\left(A_M\partial_NA_P+\frac{2}{3}A_MA_NA_P\right).
\label{CS}
\end{align} 
By substituting (\ref{decomposition of gauge fields}) into (\ref{CS}),
we make the dimensional reduction of the CS theory as follows:
\begin{align}
S_{CS}
&\rightarrow \frac{kR}{2}\int_{\Sigma_g} d^2x \:\epsilon^{\mu\nu}\mbox{Tr}(
A_{\mu}\partial_{\nu}\phi+\phi\partial_{\mu}A_{\nu}+\phi[A_{\mu},A_{\nu}])
\nonumber\\
&=\frac{kR}{2}\int_{\Sigma_g} d^2x \:\epsilon^{\mu\nu}\mbox{Tr}\left(\phi f_{\mu\nu}
+\frac{1}{2}b_{\mu\nu}\phi^2\right)
\nonumber\\
&=\frac{kR}{2}\int_{\Sigma_g} d^2x \mbox{Tr}\left(\epsilon^{\mu\nu}\phi f_{\mu\nu}
-\frac{2\pi Rp}{V}\sqrt{g}\phi^2 \right),
\label{dimensional reduction}
\end{align}
where $f_{\mu\nu}=\partial_{\mu}a_{\nu}-\partial_{\nu}a_{\mu}
+[a_{\mu},a_{\nu}]$ and
we have used (\ref{curvature 2-form}) in the last equality.
The last line in (\ref{dimensional reduction}) takes the form of 
the BF theory with the mass term: the first term is the action of 
the BF theory on $\Sigma_g$, where $\phi$ corresponds to $B$,
and the second term is a mass term for $\phi$.
Note that $k$ in (\ref{CS}) must be an integer while such restriction
is no longer imposed on $k$ in 
(\ref{dimensional reduction}).
If we integrate $\phi$ out, we obtain the two-dimensional Yang-Mills 
on $\Sigma_g$,
\begin{align}
S_{YM}=\frac{1}{4g_{YM}^2}\int_{\Sigma_g} d^2x \sqrt{g} \:
\mbox{Tr}(f_{\mu\nu}f^{\mu\nu}),
\end{align}
where the coupling constant is given by
$g_{YM}^2=\frac{2\pi p}{kV}$.


The equations of motion for the BF theory with the mass term are
\begin{align}
&f_{\mu\nu}+b_{\mu\nu}\phi=0, \nonumber\\
&D_{\mu}\phi=0.
\end{align}
These equations are solved in the gauge in which $\phi$ is diagonal as
follows:
\begin{align}
&\hat{a}_{\mu}=-b_{\mu}\hat{\phi}, \nonumber\\
&\hat{\phi}=-\frac{i}{pR}
\mbox{diag}(\cdots,\underbrace{n_{s-1},\cdots,n_{s-1}}_{N_{s-1}},
\underbrace{n_{s},\cdots,n_{s}}_{N_{s}},
\underbrace{n_{s+1},\cdots,n_{s+1}}_{N_{s+1}},\cdots),
\label{background}
\end{align}
where $s$ label the (diagonal) 
blocks, $n_s$ must be constant integers due to
Dirac's quantization condition for the monopole charge and $\sum_s N_s=M$.

In the following, we show that we obtain the $U(N)$ 
CS theory on the total space from the $U(N\times\infty)$ BF theory with the 
mass term on the base space through the following procedure:
we choose a certain background of the $U(N\times\infty)$ 
BF theory with the mass term,
expand the theory around the background and impose a periodicity condition. 
The background is given by (\ref{background}) with 
$s$ running from $-\infty$ to $\infty$, $n_s=ps$ and $N_s=N$. 
We decompose the fields into the backgrounds and the fluctuations as 
\begin{align}
a_{\mu}&\rightarrow \hat{a}_{\mu}+a_{\mu}, \nonumber\\
\phi&\rightarrow \hat{\phi}+\phi.
\label{background and fluctuation}
\end{align}
We label the (off-diagonal) blocks by $(s,t)$ and 
impose the periodicity (orbifolding) condition on the fluctuations:
\begin{align}
&a_{\mu}^{(s+1,t+1)}=a_{\mu}^{(s,t)}
\equiv a_{\mu}^{(s-t)}, \nonumber\\
&\phi^{(s+1,t+1)}=\phi^{(s,t)}
\equiv \phi^{(s-t)}.
\label{periodicity condition}
\end{align}
The fluctuations are gauge-transformed from
$U$ to $U'$ as \cite{IIST} 
\begin{align}
&{a'}_{\!\mu}^{(s-t)}
=e^{-\frac{i}{R}(s-t)v}a_{\mu}^{(s-t)}, \nonumber\\
&{\phi'}^{(s-t)}
=e^{-\frac{i}{R}(s-t)v}\phi^{(s-t)}.
\label{gauge transformation of fluctuations}
\end{align}
We make the Fourier transformation for the fluctuations
on each patch to construct the gauge fields
on the total space from the fields on the base space:
\begin{align}
A_{\mu}(x,y)&=\sum_{w}(a_{\mu}^{(w)}(x)
+b_{\mu}(x)\phi^{(w)}(x))
e^{-i\frac{w}{R}y}, \nonumber\\
A_y(x,y)&=\sum_{w}\phi^{(w)}(x)
e^{-i\frac{w}{R}y}.
\label{Fourier transformation}
\end{align}
We see from (\ref{gauge transformation of fluctuations}) that
the lefthand sides in the above equations indeed transform from a patch to 
another patch as the gauge fields on the total space.
We substitute (\ref{background and fluctuation}) into the $U(N\times\infty)$
BF theory with the mass term
and use (\ref{Fourier transformation}):
\begin{align}
S_{BF}&=\frac{kR}{2}\int_{\Sigma_g} d^2x \:\epsilon^{\mu\nu}\mbox{Tr}\left(\phi f_{\mu\nu}
+\frac{1}{2}b_{\mu\nu}\phi^2\right) \nonumber\\
&=\frac{kR}{2}\int_{\Sigma_g} d^2x \:\epsilon^{\mu\nu} \mbox{tr}\left[
\sum_{s,t}\left((a_{\mu}+b_{\mu}\phi)^{(s,t)}
\partial_{\nu}\phi^{(t,s)}
+\phi^{(s,t)}
\partial_{\mu}(a_{\nu}+b_{\nu}\phi)^{(t,s)} \right.\right.\nonumber\\
&\left.\left.+i\frac{t-s}{R}(a_{\mu}+b_{\mu}\phi)^{(s,t)}
(a_{\nu}+b_{\nu}\phi)^{(t,s)}\right)
+2\sum_{s,t,u}\phi^{(s,t)}(a_{\mu}+b_{\mu}\phi)^{(t,u)}
(a_{\nu}+b_{\nu}\phi)^{(u,s)} \right] \nonumber\\
&=\sum_{s}\frac{k}{4\pi}\int_{\cal M} d^3z \:\epsilon^{MNP}\mbox{tr}
\left(A_{M}\partial_NA_P+\frac{2}{3}A_MA_NA_P \right),
\end{align}
where we have ignored a constant term.
By dividing an overall factor $\sum_s$ in the last line to extract a single
period, we obtain the CS theory on the total space.
In the above procedure, we obtain the theory around the trivial vacuum of 
CS theory. When $g\neq 0$ or $p \neq 1$, the 
CS theory on the total space has nontrivial vacua.
We can obtain the theories around the nontrivial vacua of CS theory from
BF theory on the base space by extending the above procedure 
in a straightforward way. (See \citen{IIST}.) 

\section{BF theory with the mass term vs matrix model }
\setcounter{equation}{0}
For the case of $g=0$, 
the base space is $S^2$ and the total space is the lens space $S^3/\Z_p$.
In this section, we show that the BF theory with the mass term on $S^2$
is realized in the matrix model that is its dimensional reduction to zero
dimensions. Combining this result with the result in the
previous section, we find that the CS theory on $S^3/\Z_p$ is realized in
the matrix model.

One needs two patches to describe the lens space $S^3/\Z_p$: the patch I
is specified by $0\leq \theta < \pi$ and the patch II is specified by
$0<\theta\leq\pi$.
We adopt the following metric for $S^3/\Z_p$:
\begin{align}
ds^2=\frac{1}{\mu^2}\left(d\theta^2+\sin^2\theta d\varphi^2
+\left(\frac{1}{p}d(\psi\pm\varphi)+(\cos\theta\mp 1) d\varphi\right)^2\right),
\label{metric of lens space}
\end{align}
where $0\leq\theta\leq\pi$, $0\leq\varphi\leq 2\pi$ and $0\leq \psi\leq 4\pi$.
The upper sign is taken in the patch I
while the lower sign in the patch II.
The radius of $S^3/\Z_p$ is $2/\mu$ and that of $S^2$ is $1/\mu$.
$R$ is given by $\frac{2}{p\mu}$ and $b_{\theta}=0, \;\; 
b_{\varphi}=\frac{1}{\mu}(\cos\theta\mp 1)$. Note that for $p=1$, 
(\ref{metric of lens space}) takes the well-known form of the metric of $S^3$:
\begin{align}
ds^2=\frac{1}{\mu^2}\left(d\theta^2+\sin^2\theta d\varphi^2
+(d\psi+\cos\theta d\varphi)^2\right).
\label{metric of S3}
\end{align}

It is convenient 
to rewrite the BF theory with the mass term on $S^2$ 
by using the three-dimensional
flat-space notation. We define vector fields in terms of the gauge
fields and the Higgs field on $S^2$:
\begin{align}
\vec{y}=-i\left(\frac{1}{\mu}\phi \vec{e}_r + a_{\theta}\vec{e}_{\varphi}
-\frac{1}{\sin\theta}a_{\phi}\vec{e}_{\theta}\right),
\end{align}
where 
$\vec{e}_r=(\sin\theta\cos\varphi,\sin\theta\sin\varphi,\cos\theta)$ and
$\vec{e}_{\theta}=\frac{\partial \vec{e}_r}{\partial \theta},\;\;
\vec{e}_{\varphi}=\frac{1}{\sin\theta}
\frac{\partial \vec{e}_r}{\partial \varphi}$.
We also introduce the angular momentum operator in three-dimensional flat space,\begin{align}
\vec{L}^{(0)}=-i\vec{e}_{\phi}\partial_{\theta}
+i\frac{1}{\sin\theta}\vec{e}_{\theta}\partial_{\phi}.
\label{angular momentum operator in monopole background}
\end{align}
The BF theory with the mass term on $S^2$ takes the form
\begin{align}
S_{BF}=\frac{2k}{p\mu}\int d\theta d\varphi
\mbox{Tr}\left(-\frac{1}{2\mu}\sin\theta\phi^2
+\phi (\partial_{\theta}a_{\varphi}-\partial_{\varphi}a_{\theta}
+[a_{\theta},a_{\varphi}])\right).
\end{align}
It is rewritten in terms of $y_i$ and $L_i^{(0)}$ as 
\begin{align}
S_{BF}=-\frac{1}{g_{BF}^2}
\int \frac{d\Omega_2}{4\pi} \mbox{Tr}\left\{ y_i\left(\frac{1}{2}y_i
+\frac{i}{2}\epsilon_{ijk}L^{(0)}_jy_k
+\frac{i}{6}\epsilon_{ijk}[y_j,y_k] \right)\right\},
\label{rewritten BF}
\end{align}
where $g_{BF}^2=\frac{p}{8\pi k}$.
By making a replacement $y_i\rightarrow \hat{y}_i+y_i$, where $\hat{y}_i$
denote the background for $y_i$ corresponding to (\ref{background}), we
expand the theory around (\ref{background}):
\begin{align}
S_{BF}=-\frac{1}{g_{BF}^2}
\int \frac{d\Omega_2}{4\pi} \sum_{s,t}
\mbox{tr}\left\{ y_i^{(s,t)}\left(\frac{1}{2}y_i^{(t,s)}
+\frac{i}{2}\epsilon_{ijk}L^{(q_{ts})}_jy_k^{(t,s)}
+\frac{i}{6}\epsilon_{ijk}[y_j,y_k]^{(t,s)} \right)\right\},
\label{BF expanded around background}
\end{align}
where $q_{ts}=(n_t-n_s)/2$ and we have ignored a constant term.
$\vec{L}^{(q)}$ is the angular momentum operator in the presence of a monopole
with the magnetic charge $q$ at the origin, which takes 
the form \cite{Wu:1976ge}
\begin{align}
\vec{L}^{(q)}=\vec{L}^{(0)}-q\frac{\cos\theta\mp 1}{\sin\theta}\vec{e}_{\theta}
-q\vec{e}_r,
\end{align}
where the upper sign is taken in the patch I and the lower sign
in the patch II.

We obtain a matrix model by dropping the derivatives in (\ref{rewritten BF}):
\begin{align}
S_{mm}=-\frac{1}{g_{mm}^2}\mbox{Tr}\left\{ X_i\left( \frac{1}{2}X_i
+\frac{i}{6}\epsilon_{ijk}[X_j,X_k]\right) \right\},
\label{matrix model}
\end{align}
where $X_i$ are $M \times M$ hermitian matrices.
Note that this matrix model is nothing but 
the Dijkgraaf-Vafa matrix model which
reproduces the effective superpotential 
of ${\cal N}=1^*$ theory \cite{Kazakov:1998ji,Dijkgraaf:2002dh,Dorey:2002tj}.
We will find a relationship between
the BF theory with the mass term on $S^2$ and the above matrix model.
There the Higgs branch of ${\cal N}=1^*$ theory, which
is classified by an irreducible decomposition of $SU(2)$ representation,
plays an important role in the equivalence.

A general solution to the equations of motion of the matrix model is given by
\begin{align}
\hat{X}_i=L_i,
\end{align}
where $L_i$ are the representation matrices of the $SU(2)$ generators
which are in general
reducible, and are decomposed into irreducible representations:
\begin{align}
L_i=
 \begin{pmatrix}
  \rotatebox[origin=tl]{-35}
  {$\cdots \;\;\;
  \overbrace{\rotatebox[origin=c]{35}{$L_{i}^{[j_{s-1}]}$} \;
  \cdots \;
  \rotatebox[origin=c]{35}{$L_{i}^{[j_{s-1}]}$}}^{\rotatebox{35}{$N_{s-1}$}}
  \;\;\;
  \overbrace{\rotatebox[origin=c]{35}{$L_i^{[j_{s}]}$} \;
  \cdots \;
  \rotatebox[origin=c]{35}{$L_i^{[j_{s}]}$}}^{\rotatebox{35}{$N_{s}$}}
  \;\;\;
  \overbrace{\rotatebox[origin=c]{35}{$L_{i}^{[j_{s+1}]}$} \;
  \cdots \;
  \rotatebox[origin=c]{35}{$L_{i}^{[j_{s+1}]}$}}^{\rotatebox{35}{$N_{s+1}$}}
  \;\;\;\cdots $}
 \end{pmatrix},
 \label{matrix background}
\end{align}
where $L_i^{[j]}$ are the spin $j$ representation matrices of $SU(2)$ and 
$M=\sum_s N_s(2j_s+1)$. 
By making a replacement $X_i\rightarrow \hat{X}_i+X_i$, 
we expand the theory around (\ref{matrix background}):
\begin{align}
S_{mm}=-\frac{1}{g_{mm}^2}\sum_{s,t}
\mbox{tr}\left\{ X_i^{(s,t)}\left( \frac{1}{2}X_i^{(t,s)}
+\frac{i}{2}\epsilon_{ijk}L_j\circ X_k^{(t,s)}
+\frac{i}{6}\epsilon_{ijk}[X_j,X_k]^{(t,s)}\right) \right\},
\label{matrix model expanded around background}
\end{align}
where $L_i\circ$ is defined by
\begin{align}
L_i\circ X_j^{(s,t)}=L_i^{[j_s]}X_j^{(s,t)}
-X_j^{(s,t)}L_i^{[j_t]},
\end{align}
and we have ignored a constant term.

In what follows, we show that the theory around (\ref{background}) of 
the BF theory with the mass term is equivalent to the theory around 
(\ref{matrix background}) with $2j_s+1=N_0+n_s$ of the matrix model
in the $N_0 \rightarrow \infty$ limit. For this purpose, we make a harmonic
expansion of (\ref{BF expanded around background}) and (\ref{matrix model
expanded around background}).
In (\ref{BF expanded around background}), we expand the fields in terms of
the monopole vector spherical harmonics $\tilde{Y}_{Jmqi}^{\rho}$ defined
in (\ref{monopole scalar spherical harmonics}) and 
(\ref{vector spherical harmonics}) as
\begin{align}
y_i^{(s,t)}=\sum_{\rho=0,\pm 1}\sum_{\tilde{Q}\geq |q_{st}|}
\sum_{m=-Q}^Q y_{Jm\rho}^{(s,t)}\tilde{Y}_{Jmqi}^{\rho},
\label{mode expansion of y}
\end{align}
where $Q=J+\frac{(1+\rho)\rho}{2}$ and $\tilde{Q}=J-\frac{(1-\rho)\rho}{2}$.
Substituting (\ref{mode expansion of y}) into 
(\ref{BF expanded around background}) yields
\begin{align}
S_{BF}&=-\frac{1}{g_{BF}^2}\mbox{tr}\left(
\frac{1}{2}\sum_{s,t}\rho (J+1) y_{Jm\rho}^{(s,t)\dagger}y_{Jm\rho}^{(s,t)}
\right.\nonumber\\
&\left.\qquad+\frac{i}{3}\sum_{s,t,u}
{\cal E}_{J_1m_1q_{st}\rho_1\;J_2m_2q_{tu}\rho_2\;J_3m_3q_{us}\rho_3}
y_{J_1m_1\rho_1}^{(s,t)}y_{J_2m_2\rho_2}^{(t,u)}y_{J_3m_3\rho_3}^{(u,s)}
\right),
\label{mode expansion of BF}
\end{align}
where (\ref{action of SU(2) on vector spherical harmonics}), 
(\ref{orthonormal relation for vector spherical harmonics}) and
(\ref{integral of product of three vector spherical harmonics}) were used.
In (\ref{matrix model expanded around background}), we expand the matrices
in terms of fuzzy vector spherical harmonics $\hat{Y}_{Jm(j_sj_t)i}^{\rho}$ as
\begin{align}
X_i^{(s,t)}=\sum_{\rho=0,\pm 1}\sum_{\tilde{Q}\geq |j_s-j_t|}^{j_s+j_t}
\sum_{m=-Q}^Q X_{Jm\rho}^{(s,t)}\otimes \hat{Y}_{Jm(j_sj_t)i}^{\rho}.
\label{mode expansion of X}
\end{align}
Since $j_s+j_t=N_0+\frac{n_s+n_t}{2}-1$, $N_0$ plays a role of the 
ultraviolet cutoff. Note also that $j_s-j_t=(n_s-n_t)/2=q_{st}$.
Substituting (\ref{mode expansion of X}) into (\ref{matrix model expanded
around background}) yields
\begin{align}
S_{mm}&=-\frac{N_0}{g_{mm}^2}\mbox{tr}\left(
\frac{1}{2}\sum_{s,t}\rho (J+1) X_{Jm\rho}^{(s,t)\dagger}X_{Jm\rho}^{(s,t)}
\right.\nonumber\\
&\left.\qquad+\frac{i}{3}\sum_{s,t,u}
\hat{{\cal E}}_{J_1m_1(j_sj_t)\rho_1\;J_2m_2(j_tj_u)\rho_2\;
J_3m_3(j_uj_s)\rho_3}
X_{J_1m_1\rho_1}^{(s,t)}X_{J_2m_2\rho_2}^{(t,u)}X_{J_3m_3\rho_3}^{(u,s)}
\right),
\label{mode expansion of matrix model}
\end{align}
where (\ref{action of SU(2) on vector spherical harmonics}), 
(\ref{orthonormal relation for vector spherical harmonics}) and
(\ref{integral of product of three vector spherical harmonics}) 
were again used.
In the $N_0\rightarrow \infty$ limit, the ultraviolet cutoff goes to infinity 
and $\hat{{\cal E}}_{J_1m_1(j_sj_t)\rho_1\;J_2m_2(j_tj_u)\rho_2\;
J_3m_3(j_uj_s)\rho_3}$ reduces to
${\cal E}_{J_1m_1q_{st}\rho_1\;J_2m_2q_{tu}\rho_2\;
J_3m_3q_{us}\rho_3}$ as shown in appendix. Hence, in the limit in which 
$N_0\rightarrow \infty$ and $g_{mm}\rightarrow \infty$ 
such that 
$g_{mm}^2/N_0=g_{BF}^2$, 
(\ref{mode expansion of matrix model}) agrees 
with (\ref{mode expansion of BF}) under the identification
$X_{Jm\rho}^{(s,t)}=y_{Jm\rho}^{(s,t)}$. We have proven our statement.

Combining the above result with the result in the previous section, we see
that the $U(N)$ CS theory around the trivial vacuum on $S^3/\Z_p$ 
is realized in
the matrix model as follow. In (\ref{matrix background}), we make $s$ run
from $-\infty$ to $\infty$ and put $N_s=N$ and $2j_s+1=N_0+ps$. We impose the
periodicity condition: 
$X_{Jm\rho}^{(s+1,t+1)}=X_{Jm\rho}^{(s,t)}=X_{Jm\rho}^{(s-t)}$.
We take the limit in which $N_0\rightarrow \infty$ and 
$g_{mm}\rightarrow \infty$ such that 
$g_{mm}^2/N_0=\frac{p}{8\pi k}$ and divide the overall factor $\sum_s$ out.
Thus we obtain the CS theory around the trivial vacuum on $S^3/\Z_p$.
Indeed, if we expand the gauge fields $A_i$ expressed in the local Lorentz
frame in terms of the vector spherical harmonics on $S^3$ defined
in (\ref{scalar spherical harmonics on S^3}) and 
(\ref{vector spherical harmonics}) as 
\begin{align}
A_i=\sum_{Jm\tilde{m}} A_{Jm\tilde{m}}Y_{Jm\tilde{m}i}^{\rho},
\end{align}
we obtain a harmonic expansion of the CS theory on $S^3/\Z_p$ as follows:
\begin{align}
S_{CS}&=\frac{8\pi k}{p\mu^3}\int \frac{d\Omega_3}{2\pi^2/p}
\epsilon^{ijk}\mbox{Tr}\left(\frac{1}{2}A_i\nabla_jA_k+\frac{1}{3}A_iA_jA_k
\right) \nonumber\\
&=\frac{8\pi k}{p\mu^3}
\mbox{Tr}\left(\frac{1}{2}\mu \rho (J+1)
A_{Jm\tilde{m}\rho}^{\dagger}A_{Jm\tilde{m}\rho} \right.\nonumber\\
&\left.\qquad\qquad
+\frac{1}{3}{\cal E}_{J_1m_1\tilde{m_1}\rho_1\;J_2m_2\tilde{m}_2\rho_2\;
J_3m_3\tilde{m}_3\rho_3}
A_{J_1m_1\tilde{m}_1\rho_1}A_{J_2m_2\tilde{m}_2\rho_2}
A_{J_3m_3\tilde{m}_3\rho_3} \right),
\label{mode expansion of CS}
\end{align}
where $\tilde{m},\;\tilde{m}_1,\;\tilde{m}_2,\;\tilde{m}_3$ 
are restricted to $\frac{p}{2}\Z$.
If we compare (\ref{mode expansion of CS}) with
(\ref{mode expansion of matrix model}), we see that
the relation between the modes is given by
\begin{align}
A_{Jm\tilde{m}\rho}=i\mu X_{Jm\rho}^{(2\tilde{m}/p)}.
\label{relation between modes}
\end{align}
It is now easy to obtain the CS theory around a nontrivial vacuum on $S^3/\Z_p$
from the matrix model.

\section{Description of Wilson loop in CS on $S^3$ by matrices}
\setcounter{equation}{0}
Let $z^{M}(\sigma)$ parameterize a closed loop on $S^3$, where
$M=\theta,\varphi,\psi$, $0\leq\sigma\leq 1$ and 
$z^{M}(0)=z^{M}(1)=z^{M}$.
Then we consider a Wilson loop on $S^3$, which takes the form
\begin{align}
W &=\mbox{Tr}\left[ P\exp\left(\int_0^1A_{M}(z(\sigma))
\frac{dz^{M}(\sigma)}{d\sigma}d\sigma\right)\right] \n
&=\mbox{Tr}\left[ P\exp\left(\int_0^1A_{i}(z(\sigma))
e^i_{M}(z(\sigma))
\frac{dz^{M}(\sigma)}{d\sigma}d\sigma\right)\right],
\label{Wilson loop}
\end{align}
where $e^i_{M}$ $(i=1,2,3)$ is the right-invariant 1-form on $S^3$ defined
in appendix.
We divide the loop into $n$ small bits denoted by $\Delta z_a^{M}$
$(a=1,\cdots,n)$ and take the $n\rightarrow\infty$ limit.
By definition the bits satisfy $\sum_{a=1}^n \Delta z_a^{M}=0$.
The Wilson loop (\ref{Wilson loop}) is rewritten as
\begin{align}
W &=\mbox{Tr}\left[\prod_{a=1}^n
\left(1+A_{i_a}(z+\sum_{b=1}^{a-1}\Delta z_b)
e_{M_a}^{i_a}(z+\sum_{c=1}^{a-1}\Delta z_c)\Delta z_a^{M_a}\right)
\right] \n
&=\mbox{Tr}\left[
\Big(1+A_{i_1}(z)e_{M_1}^{i_1}(z)\Delta z_1^{M_1}\Big)
\Big(1+A_{i_2}(z+\Delta z_1)e_{M_2}^{i_2}(z+\Delta z_1)\Delta z_2^{M_2}
\Big)\cdots
\right.\n 
& \qquad\;\;\;\cdots
\Big(1+A_{i_{n-1}}(z-\Delta z_{n-1}-\Delta z_{n})
e_{M_{n-1}}^{i_{n-1}}(z-\Delta z_{n-1}-\Delta z_n)\Delta z_{n-1}^{M_{n-1}}
\Big)\n
& \left.\qquad\qquad
\Big(1+A_{i_n}(z-\Delta z_n)
e_{M_n}^{i_n}(z-\Delta z_n)\Delta z_n^{M_n}\Big)
\right].
\label{Wilson loop 2}
\end{align}

We expand the gauge fields in terms of the scalar spherical harmonics on $S^3$
defined in (\ref{scalar spherical harmonics on S^3})\footnote{In this
section, we expand the vector fields in terms of the scalar harmonics to
make the discussion simpler.}:
\begin{align}
A_i(z)=\sum_{Jm\tilde{m}}Y_{Jm\tilde{m}}(z)A_{Jm\tilde{m}i}.
\end{align}
Then the gauge fields at $z+\Delta z_1$ are evaluated as 
\begin{align}
A_i(z+\Delta z_1) 
&=\sum_{Jm\tilde{m}}Y_{Jm\tilde{m}}(z+\Delta z_1)A_{Jm\tilde{m}i} \n
&=\sum_{Jm\tilde{m}}e^{\Delta z_1^{M_1}\partial_{M_1}} 
Y_{Jm\tilde{m}}(z)A_{Jm\tilde{m}i} \n
&=\sum_{Jm\tilde{m}}e^{i\mu\Delta z_1^{M_1} e_{M_1}^{i_1}(z){\cal L}_{i_1}} 
Y_{Jm\tilde{m}}(z)A_{Jm\tilde{m}i} \n
&=\sum_{Jmm_1\tilde{m}}Y_{Jm_1\tilde{m}}(z)
\langle Jm_1|e^{i\mu\Delta z_1^{M_1} e_{M_1}^{i_1}(z)J_{i_1}}|Jm\rangle
A_{Jm\tilde{m}i}.
\label{gauge field at translated point 1}
\end{align}
Here ${\cal L}_i$ are the Killing vectors that obey the $SU(2)$ algebra and
equal  $-\frac{i}{\mu}e^{M}_i\partial_{M}$, where $e^{M}_i$ 
are the inverse of $e_{M}^i$. 
$\langle Jm_1|e^{i\Delta z_1^{M_1} e_{M_1}^{i_1}(z) J_{i_1}}
|Jm\rangle$ is the matrix element of the spin $J$ representation for
an $SU(2)$ element $e^{i\Delta z_1^{M_1} e_{M_1}^{i_1}(z) J_{i_1}}$.
The gauge fields at $z+\Delta z_1+\Delta z_2$ are evaluated as 
\begin{align}
A_i(z+\Delta z_1+\Delta z_2) 
&=\sum_{Jm\tilde{m}}
e^{i\mu\Delta z_2^{M_2} e_{M_2}^{i_2}(z+\Delta z_1){\cal L}_{i_2}}
Y_{Jm\tilde{m}}(z+\Delta z_1)A_{Jm\tilde{m}i} \n
&=\sum_{Jmm_2\tilde{m}}Y_{Jm_2\tilde{m}}(z+\Delta z_1)
\langle Jm_2|e^{i\mu\Delta z_2^{M_2} e_{M_2}^{i_2}(z+\Delta z_1)J_{i_2}}
|Jm\rangle
A_{Jm\tilde{m}i} \n
&=\sum_{Jmm_1m_2\tilde{m}}Y_{Jm_1\tilde{m}}(z)
\langle Jm_1|e^{i\mu\Delta z_1^{M_1}e_{M_1}^{i_1}(z)J_{i_1}}|Jm_2\rangle \n
&\qquad\qquad\;\;\;\;\;
\times\langle Jm_2|
e^{i\mu\Delta z_2^{M_2}e_{M_2}^{i_2}(z+\Delta z_1)J_{i_2}}
|Jm\rangle
A_{Jm\tilde{m}i}.
\label{gauge field at translated point 2}
\end{align}
Similarly we can evaluate $A_{i_a}(z+\sum_{b=1}^{a-1}\Delta z_b)$ and
express the Wilson loop (\ref{Wilson loop 2}) 
in terms of the spherical harmonics at $z$.
Due to homogeneity of $S^3$, we can consider a set of Wilson loops starting
and ending at $z$ such that
$e_{M_a}^{i_a}(z+\sum_{b=1}^{a-1}\Delta z_b)\Delta z_a^{M_a}$ 
is independent of $z$.
Then we average the Wilson loops over $S^3$:
\begin{align}
\tilde{W}=\int \frac{d\Omega_3}{2\pi^2} \: W,
\end{align}
where the integration acts only on products of 
$Y_{Jm\tilde{m}}(z)$. Note that $\langle W \rangle=\langle \tilde{W} \rangle$.

Correspondingly, we can consider the Wilson loop in the matrix model:
\begin{align}
\hat{W}&=\frac{1}{TN_0}
\mbox{Tr}\left[P\exp \left(i\mu\int_0^1 X_i e_{M}^i(z(\sigma))
\frac{dz^{M}(\sigma)}{d\sigma}d\sigma\right)\right] \n
&=\frac{1}{TN_0}\mbox{Tr}\left[\prod_{a=1}^n
\left(1+i\mu X_{i_a}
e_{M_a}^{i_a}(z+\sum_{b=1}^{a-1}\Delta z_b)\Delta z_a^{M_a}\right)
\right],
\label{Wilson loop in matrix model}
\end{align}
where $T=\sum_s1$.
Here we decompose $X_i$ as $X_i \rightarrow L_i +X_i$, 
where $L_i$ are
given in (\ref{matrix background}) with $s$ running from $-\infty$ to $\infty$,
$2j_s+1=N_0+s$, $N_s=N$, $N_0\rightarrow\infty$,
and the periodicity condition $X_i^{(s,t)}=X_i^{(s-t)}$ imposed.
We rewrite 
(\ref{Wilson loop in matrix model}) as
\begin{align}
\hat{W}=\frac{1}{TN_0}\mbox{Tr}\left[\prod_{a=1}^n
\left(1+i\mu X\cdot e(z+\sum_{b=1}^{a-1}\Delta z_b)\cdot\Delta z_a\right)
e^{i\mu L\cdot e(z+\sum_{c=1}^{a-1}\Delta z_c) \cdot \Delta z_a}
\right],
\label{Wilson loop in matrix model 2}
\end{align}
where $L\cdot e \cdot \Delta x=L_i e_{M}^i \Delta z^{M}$ and so on.
We further evaluate (\ref{Wilson loop in matrix model 2}) as follows:
\begin{align}
\hat{W} &=
\frac{1}{TN_0}\mbox{Tr}\left[(1+i\mu X\cdot e(z) \cdot \Delta z_1)
e^{i\mu L \cdot e(z) \Delta z_1}(1+i\mu X \cdot e(z+\Delta z_1)\cdot\Delta z_2)
e^{-i\mu L \cdot e(z) \cdot \Delta z_1} \right.\n
& \qquad
e^{i\mu L \cdot e(z) \cdot \Delta z_1}
e^{i\mu L \cdot e(z+\Delta z_1)\cdot\Delta z_2}
(1+i\mu X \cdot e(z+\Delta z_1+\Delta z_2)\cdot\Delta z_3) \n
& \qquad \times e^{-i\mu L \cdot e(z+\Delta z_1) \cdot \Delta z_2}
e^{-i\mu L \cdot e(z) \cdot \Delta z_1} \n
& \qquad \cdots \n
& \qquad
e^{i\mu L \cdot e(z) \cdot \Delta z_1}\cdots 
e^{i\mu L \cdot e(z-\Delta z_{n-1}-\Delta z_n) \cdot \Delta z_{n-1}}
(1+i\mu X \cdot e(z-\Delta z_n)\cdot\Delta z_n) \n
& \qquad \times 
e^{-i\mu L \cdot e(z-\Delta z_{n-1}-\Delta z_n) \cdot \Delta z_{n-1}}\cdots
e^{-i\mu L \cdot e(z) \cdot \Delta z_1}  \n
& \left. \qquad
e^{i\mu L \cdot e(z) \cdot \Delta z_1}\cdots 
e^{i\mu L \cdot e(z-\Delta z_n) \cdot \Delta z_n} \right].
\label{Wilson loop in matrix model 3}
\end{align}
Note that the factor $e^{i\mu L \cdot e(z) \cdot \Delta z_1}\cdots 
e^{i\mu L \cdot e(z-\Delta z_n) \cdot \Delta z_n}$ appearing in the last
line of (\ref{Wilson loop in matrix model 3}) equals the identity if 
it is invariant under any deformation of the loop.
In order to see this invariance, we consider two paths which start at $z$ and
end at $z+\Delta z+\Delta z'$: 
(1) $z\rightarrow z+\Delta z \rightarrow z+\Delta z +\Delta z'$ and
(2) $z\rightarrow z+\Delta z' \rightarrow z+\Delta z +\Delta z'$.
We associate $e^{i\mu L \cdot e(z) \cdot \Delta z}
e^{i\mu L \cdot e(z+\Delta z)\cdot\Delta z'}$ and
$e^{i\mu L \cdot e(z) \cdot \Delta z'}
e^{i\mu L \cdot e(z+\Delta z')\cdot\Delta z}$ with (1) and (2), respectively.
The difference between these quantities is evaluated 
up to ${\cal O}((\Delta z)^3)$ as
\begin{align}
&e^{i\mu L \cdot e(z) \cdot \Delta z}
e^{i\mu L \cdot e(z+\Delta z)\cdot\Delta z'}
-e^{i\mu L \cdot e(z) \cdot \Delta z'}
e^{i\mu L \cdot e(z+\Delta z')\cdot\Delta z} \n
&=i\mu(\partial_{M}e_{M'}^i(z)-\partial_{M'}e_{M}^i(z)
-\mu f_{ijj'}e_{M}^j(z) e_{M'}^{j'}(z)) L_i 
\Delta z^{M} \Delta {z'}^{M'}.
\end{align}
This vanishes thanks to the Maurer-Cartan equation.
This fact indicates that the factor
$e^{-i\mu L \cdot e(z) \cdot \Delta z_1}\cdots 
e^{-i\mu L \cdot e(z-\Delta z_n) \cdot \Delta z_n}$ associated with 
a closed path equals the identity.
Eventually, the Wilson loop (\ref{Wilson loop in matrix model})
takes the form
\begin{align}
\hat{W}=&\frac{1}{TN_0}\mbox{Tr}\left[\prod_{a=1}^n
e^{i\mu L\cdot e(z) \cdot \Delta z_1} \cdots
e^{i\mu L\cdot e(z+\sum_{b=1}^{a-2}\Delta x_b)\cdot \Delta z_{a-1}} \right.
\n
& \times \left(1+i\mu X\cdot e(z+\sum_{c=1}^{a-1}\Delta z_c)\cdot\Delta
 z_a\right) 
\left. 
e^{-i\mu L\cdot e(z+\sum_{d=1}^{a-2}\Delta z_d)\cdot \Delta z_{a-1}} \cdots
e^{-i\mu L\cdot e(z) \cdot \Delta z_1}
\right].
\label{Wilson loop in matrix model 4}
\end{align}

We expand the $(s,t)$ block of $X_i$ in terms of the fuzzy scalar
spherical harmonics as
\begin{align}
X_i^{(s,t)}=\sum_{Jm}\hat{Y}_{Jm}^{(j_sj_t)}X_{Jmi}^{(s,t)}.
\end{align}
We evaluate an expression appearing in the Wilson loop:
\begin{align}
e^{i\mu L\cdot e(z) \cdot \Delta z_1}X_i^{(s,t)}
e^{-i\mu L\cdot e(z) \cdot \Delta z_1}
&=\sum_{Jm}e^{i\mu L\cdot e(z) \cdot \Delta z_1}\hat{Y}_{Jm}^{(j_sj_t)}
e^{-i\mu L\cdot e(z) \cdot \Delta z_1} X_{Jmi}^{(s,t)} \n
&=\sum_{Jmm_1}\hat{Y}_{Jm_1}^{(j_sj_t)}
\langle Jm_1|e^{i\mu\Delta z_1 \cdot e(z) \cdot J}|Jm\rangle X_{Jmi}^{(s,t)}. 
\label{matrix at translated point 1}
\end{align}
We evaluate another expression:
\begin{align}
&e^{i\mu L\cdot e(z) \cdot \Delta z_1}
e^{i\mu L\cdot e(z+\Delta z_1) \cdot \Delta z_2}
X_i^{(s,t)}
e^{-i\mu L\cdot e(z+\Delta z_1) \cdot \Delta z_2}
e^{-i\mu L\cdot e(z) \cdot \Delta z_1} \n
&=\sum_{Jmm_2}e^{i\mu L\cdot e(z) \cdot \Delta z_1}\hat{Y}_{Jm_2}^{(j_sj_t)}
e^{-i\mu L\cdot e(z) \cdot \Delta z_1} 
\langle Jm_2|e^{i\mu\Delta z_2 \cdot e(z+\Delta z_1) \cdot J}|Jm\rangle
X_{Jma}^{(s,t)} \n
&=\sum_{Jmm_1m_2\tilde{m}}\hat{Y}_{Jm_1}^{(j_sj_t)}
\langle Jm_1|e^{i\mu\Delta z_1 \cdot e(z) \cdot J}|Jm_2\rangle \n
&\qquad\qquad\;\;\;\;\;
\times\langle Jm_2|e^{i\mu\Delta z_2 \cdot e(z+\Delta z_1) \cdot J}|Jm\rangle
X_{Jmi}^{(s,t)}.
\label{matrix at translated point 2}
\end{align}
In this way, we can express the Wilson loop 
(\ref{Wilson loop in matrix model 4}) in terms of 
the fuzzy spherical harmonics.

Using the formulae in appendix, we can easily show that
for arbitrary $K$ 
in the $N_0\rightarrow\infty$ limit
\begin{align}
\frac{1}{TN_0}\mbox{Tr}
(\hat{Y}_{J_1m_1(j_{s_1}j_{t_1})}\cdots\hat{Y}_{J_Km_K(j_{s_K}j_{t_K})})
\rightarrow 
\int \frac{d\Omega_3}{2\pi^2}Y_{J_1m_1\tilde{m}_1}(x)\cdots
Y_{J_Km_K\tilde{m}_K}(x),
\end{align}
where $t_{\alpha}=s_{\alpha+1}$, $t_K=s_1$ and
$j_{s_{\alpha}}-j_{t_{\alpha}}=(s_{\alpha}-t_{\alpha})/2
=\tilde{m}_{\alpha}$.
Then, by comparing (\ref{gauge field at translated point 1}) with 
(\ref{matrix at translated point 1}) and 
(\ref{gauge field at translated point 2}) with 
(\ref{matrix at translated point 2}) and using (\ref{relation between modes}), 
we conclude that
\begin{align}
\hat{W} \rightarrow \tilde{W}
\end{align}
in the $N_0\rightarrow\infty$ limit.

\section{Summary and discussion}
\setcounter{equation}{0}
In this paper, we first found the relationship between
the CS theory on the total space of the $U(1)$ bundle over $\Sigma_g$ and
the BF theory with the mass term on $\Sigma_g$.
We showed that the former with the $U(N)$ gauge symmetry
is obtained by expanding the latter with $U(N\times\infty)$ gauge symmetry
around the background (\ref{background}) with $s$ running from $-\infty$
to $\infty$, $n_s=ps$ and $N_s=N$ and the periodicity condition
(\ref{periodicity condition}) imposed.
We next restricted ourselves to the case of $g=0$ and found the 
relationship between the BF theory with the mass term and the matrix model
(\ref{matrix model}).
We showed that the theory around each background of the former is equivalent
to the theory around a certain background of the latter.
By combining the above two findings, we found that the CS theory on $S^3/\Z_p$
is equivalent to the theory around the background (\ref{matrix background})
of the matrix model in the $N_0\rightarrow\infty$ limit, 
where $s$ runs from $-\infty$
to $\infty$, $2j_s+1=N_0+ps$, $N_s=N$
and the periodicity condition is imposed.
We also constructed the Wilson loops in the matrix model that correspond to
those in the CS theory on $S^3$.

It is important to see whether BF theory with mass term on $\Sigma_g$ with
$g\neq 0$ is realized in a matrix model and the CS theory on the $U(1)$ bundle
over $\Sigma_g$ with $g\neq 0$ is further realized in the matrix model.

The equivalences we found are classical ones. It is not obvious that
the equivalences hold at the quantum level\footnote{For the studies of 
quantum corrections in the related models, 
see \citen{Iso:2001mg,Imai:2003vr,
Azuma:2004zq,Kaneko:2007ui} and references therein.}.
We expect from the following discussion that this is the case. 
The phenomenon which induces the mass term for the Higgs field via
the compactification on the non-trivial $U(1)$ fiber bundle
is equivalent to the moduli stabilization by flux 
(see eg. \citen{Douglas:2006es,Blumenhagen:2006ci} and references therein) and
the $\Omega$-background in \citen{Nekrasov:2003rj}.
These background fluxes and compactifications lift up the flat directions
of the Higgs fields thanks to the induced mass term.
If this moduli is also stabilized at the quantum level
and the localization mechanism works,
we can evaluate exactly the partition function
by counting the isolated (BPS) vacua 
to show that the relationships 
among the CS theory, the BF theory and the matrix
model hold at the quantum level.
In the context of the large $N$ reduced model, the moduli stabilization 
means that we need no quenching prescription.
Indeed, the works \citen{Steinacker:2003sd,Steinacker:2007iq} suggest 
that the moduli stabilization and the localization mechanism work in 
the relationship between the BF theory and the matrix model at
the quantum level.
We would like to discuss this point in the near future.

In (\ref{BF expanded around background}), we ignored the constant term,
which depend on the background and takes the form
\begin{align}
S_{BF}^{(b.g.)}=-\frac{\pi k}{p}\sum_{s}N_s n_s^2.
\label{BF background}
\end{align}
On the other hand, in (\ref{matrix model expanded around background}),
we ignored the following constant term:
\begin{align}
S_{mm}^{(b.g)}&=-\frac{4\pi k}{3pN_0}\mbox{Tr}(L_i^2) \nonumber\\
&=-\frac{4\pi k}{3pN_0}\sum_sN_s(2j_s+1)j_s(j_s+1) \nonumber\\
&=-\frac{\pi k}{p}\left(\frac{1}{3}MN_0^2+N_0\sum_sN_sn_s+\sum_sN_sn_s^2
-\frac{1}{3}M+{\cal O}\left(\frac{1}{N_0}\right)\right).
\label{matrix model background}
\end{align}
We see that (\ref{BF background}) and
(\ref{matrix model background}) coincide in the $N_0\rightarrow \infty$ limit
up to a constant 
independent of the background
as far as we fix the first Chern class $\sum_sN_sn_s$ 
of the background on $S^2$.
This fact would be relevant when we sum up over the backgrounds 
in the path integral.

Once we verify the relationship between the CS theory on $S^3/\Z_p$ 
and the matrix model at the quantum level,
we hope that using the Wilson loops in the
matrix model constructed in section 4 we can compute the knot invariants.
We expect wide application of the Wilson loops constructed in section 4, since
they are independent of the theory
we consider. For instance, it was suggested in \citen{ISTT} that
${\cal N}=4$ super Yang Mills  
on $R\times S^3$ is realized in the plane wave
matrix model in the same manner as the CS theory on $S^3$ is realized
in the matrix model.
Namely, we can construct the Wilson loops in the plane wave matrix model
that correspond to those in ${\cal N}=4$ super Yang Mills on $R\times S^3$.
In particular, by including the six scalars in the Wilson loops,
we can construct the half-BPS Wilson loops 
on $R\times S^3$ \cite{Drukker:2007qr} 
in term of the matrices.


\section*{Acknowledgements}
K.O. would like to thank T.~Higaki, K.~Takenaga and S.~Watamura
for useful discussions and comments.
A.T. would like to thank Humboldt University for hospitality, where part of
this work was done.
The work of G.I. is supported in part by the JSPS Research Fellowship for Young
Scientists.
The work of K.O. and A.T. is supported in part by Grant-in-Aid for Scientific
Research (Nos.19740120 and 19540294) from the Ministry 
of Education, Culture, Sports, Science and Technology, respectively.

\appendix

\section{Spherical harmonics}
\setcounter{equation}{0}
\renewcommand{\theequation}{A.\arabic{equation}}
In this appendix, we review the properties of the spherical harmonics
on $S^3$, $S^2$ and fuzzy sphere summarized in \citen{ISTT,ITT}, and add
some new formulae.
We regard $S^3$ as the $SU(2)$ group manifold.
We parameterize an element of $SU(2)$ in terms of the Euler angles as
\begin{align}
g=e^{-i\varphi J_3}e^{-i\theta J_2}
e^{-i\psi J_3},
\end{align}
where $J_i$ satisfy $[J_i,J_j]=i\epsilon_{ijk}J_k$. The isometry of $S^3$
is $SO(4)=SU(2)\times SU(2)$, and these two $SU(2)$'s act on $g$ from left
and right, respectively.
We construct the right-invariant 1-forms:
\begin{align}
dgg^{-1}=-i\mu e^i J_i,
\end{align}
where $2/\mu$ corresponds to the radius of $S^3$.
They are explicitly given by
\begin{align}
e^1 &= \frac{1}{\mu}(-\sin\varphi d\theta +\sin\theta\cos\varphi d\psi), 
\nonumber\\
e^2 &= \frac{1}{\mu}(\cos\varphi d\theta +\sin\theta\sin\varphi d\psi),
\nonumber\\
e^3 &=\frac{1}{\mu}(d\varphi +\cos\theta d\psi),
\end{align}
and satisfy the Maurer-Cartan equation
\begin{align}
de^i-\frac{\mu}{2}\epsilon_{ijk}e^j\wedge e^k=0.
\end{align}
The metric constructed from $e^i_M\;\;(M=\theta,\varphi,\psi)$ 
agrees with (\ref{metric of S3}).
The Killing vectors dual to $e^i$ are given by
\begin{align}
{\cal L}_i=-\frac{i}{\mu}e_i^{M}\partial_{M},
\end{align}
where $e_i^M$ are inverse of $e^i_M$. The explicit form of the Killing
vectors is
\begin{align}
{\cal L}_1&=-i\left(-\sin\varphi\partial_{\theta}
-\cot\theta\cos\varphi\partial_{\varphi}
+\frac{\cos\varphi}{\sin\theta}\partial_{\psi}\right), \nonumber\\
{\cal L}_2&=-i\left(\cos\varphi\partial_{\theta}
-\cot\theta\sin\varphi\partial_{\varphi}
+\frac{\sin\varphi}{\sin\theta}\partial_{\psi}\right), \nonumber\\
{\cal L}_3&=-i\partial_{\varphi}.
\label{Killing vectors}
\end{align}
${\cal L}_i$ satisfy the $SU(2)$ algebra.

In the following expressions, the upper sign is taken in the patch I 
($0\leq \theta <\pi$) and 
the lower sign in the patch II ($0< \theta \leq \pi$).
Since $S^3$ is a $U(1)$ bundle over $S^2$ and $y=(\psi\pm\varphi)/\mu$, 
the angular momentum operator in the monopole background with the monopole
charge $q$ is obtained by making a replacement in
(\ref{Killing vectors}):
\begin{align}
\frac{1}{\mu}\partial_y \rightarrow -iq.
\end{align}
The result is
\begin{align}
L_1^{(q)}&=i\left(\sin\varphi\partial_{\theta}
+\cot\theta\cos\varphi \partial_{\varphi}\right)
-q\frac{1\mp \cos\theta}{\sin\theta}\cos\varphi, \nonumber\\
L_2^{(q)}&=i\left(-\cos\varphi\partial_{\theta}
+\cot\theta\sin\varphi\partial_{\varphi}\right)
-q\frac{1\mp \cos\theta}{\sin\theta}\sin\varphi, \nonumber\\
L_3^{(q)}&=-i\partial_{\varphi}\mp q,
\end{align}
which satisfy the $SU(2)$ algebra and
agree with (\ref{angular momentum operator in monopole background}).

The scalar spherical harmonics on $S^3$ are given by
\begin{align}
Y_{Jm\tilde{m}}(\Omega_3)
=(-1)^{J-\tilde{m}}\sqrt{2J+1}\langle J\:-\tilde{m}|g^{-1}|Jm\rangle.
\label{scalar spherical harmonics on S^3}
\end{align}
The monopole scalar spherical harmonics \cite{Wu:1976ge} 
are expressed in terms of 
the scalar spherical harmonics on $S^3$:
\begin{align}
\tilde{Y}_{Jmq}(\Omega_2)=
e^{-iq(\psi\pm\varphi)}Y_{Jmq}(\Omega_3). 
\label{monopole scalar spherical harmonics}
\end{align}
The fuzzy scalar spherical harmonics are given by
\begin{align}
\hat{Y}_{Jm}^{(jj')}=\sqrt{N_0}\sum_{r,r'}(-1)^{-j+r'}C^{Jm}_{jr \; j'-r'}
|jr\rangle\langle j'r'|.
\label{fuzzy scalar spherical harmonics}
\end{align}
These spherical harmonics possess the following properties.

\noindent
\underline{Basis of $SU(2)$ algebra}
\begin{align}
{\cal L}_{\pm}Y_{Jm\tilde{m}}
&=\sqrt{(J\mp m)(J\pm m+1)}Y_{Jm\tilde{m}}, \nonumber\\
{\cal L}_3Y_{Jm\tilde{m}}&=m Y_{Jm\tilde{m}}, \nonumber\\
L_{\pm}^{(q)}\tilde{Y}_{Jmq}
&=\sqrt{(J\mp m)(J\pm m+1)}\tilde{Y}_{Jmq}, \nonumber\\
L_3^{(q)}\tilde{Y}_{Jmq}&=m\tilde{Y}_{Jmq}, \nonumber\\
L_{\pm}\circ \hat{Y}_{Jm}^{(jj')}
&=\sqrt{(J\mp m)(J\pm m+1)}\tilde{Y}_{Jm}^{(jj')}, \nonumber\\
L_3\circ\hat{Y}_{Jm}^{(jj')}&=m\tilde{Y}_{Jm}^{(jj')},
\end{align}
where $L_i\circ$ are defined by 
$L_i\circ \hat{Y}_{Jm}^{(jj')}
=L_i^{[j]}\hat{Y}_{Jm}^{(jj')}-\hat{Y}_{Jm}^{(jj')}L_i^{[j']}$ and 
satisfy the $SU(2)$ algebra.

\noindent
\underline{Complex conjugate}
\begin{align}
&(Y_{Jm\tilde{m}})^{\dagger}=(-1)^{m-\tilde{m}}Y_{J-m-\tilde{m}},\;\;\;
(\tilde{Y}_{Jmq})^{\dagger}=(-1)^{m-q}\tilde{Y}_{J-m-q},\n
&(\hat{Y}_{Jm}^{(jj')})^{\dagger}=(-1)^{m-(j-j')}\hat{Y}_{J-m}^{(j'j)}.
\end{align}
\underline{Orthonormal relation}
\begin{align}
&\int \frac{d\Omega_3}{2\pi^2}(Y_{J'm'\tilde{m}'})^{\dagger}Y_{Jm\tilde{m}}
=\delta_{JJ'}\delta_{mm'}\delta_{\tilde{m}\tilde{m}'}, \nonumber\\
&\int \frac{d\Omega_2}{4\pi}(Y_{J'm'q})^{\dagger}Y_{Jmq}
=\delta_{JJ'}\delta_{mm'}, \nonumber\\
&\frac{1}{N_0}\mbox{tr}((\hat{Y}_{J'm'}^{(jj')})^{\dagger}\hat{Y}_{Jm}^{(jj')})
=\delta_{JJ'}\delta_{mm'}.
\end{align}
\underline{Integral of the product of three harmonics}
\begin{align}
&\int \frac{d\Omega_3}{2\pi^2}(Y_{J_1m_1\tilde{m}_1})^{\dagger}
Y_{J_2m_2\tilde{m}_2}Y_{J_3m_3\tilde{m}_3}
\n
&=\sqrt{\frac{(2J_2+1)(2J_3+1)}{2J_1+1}} 
C^{J_1m_1}_{J_2m_2\;J_3m_3}C^{J_1\tilde{m}_1}_{J_2\tilde{m}_2\;J_3\tilde{m}_3}
\equiv{\cal C}^{J_1m_1\tilde{m}_1}_{J_2m_2\tilde{m}_2\;J_3m_3\tilde{m}_3},
\nonumber \\
&\int \frac{d\Omega_2}{4\pi}(\tilde{Y}_{J_1m_1q_1})^{\dagger}
\tilde{Y}_{J_2m_2q_2}\tilde{Y}_{J_3m_3q_3}
={\cal C}^{J_1m_1q_1}_{J_2m_2q_2\;J_3m_3q_3},
\nonumber \\
&\frac{1}{N_0}\mbox{tr}((\hat{Y}_{J_1m_1}^{(j'j)})^{\dagger}
\hat{Y}_{J_2m_2}^{(j'j'')}\hat{Y}_{J_3m_3}^{(j''j)})
\n
&=(-1)^{J_1+2J_3-j+j'-2j''}
\sqrt{N_0(2J_2+1)(2J_3+1)}C^{J_1m_1}_{J_2m_2\;J_3m_3}
 \begin{Bmatrix}
  J_1 & J_2 & J_3 \\
  j'' & j & j'
 \end{Bmatrix} \nonumber\\
&\equiv \hat{{\cal C}}^{J_1m_1(j'j)}_{J_2m_2(j'j'')\;J_3m_3(j''j)}.
\label{C and Chat}
\end{align}

There is a formula for the asymptotic relations between the
$6-j$ symbols and the $3-j$ symbols. If $R\gg 1$, one obtains
\begin{align}
 \begin{Bmatrix}
  a & b & c \\
  d+R & e+R & f+R
 \end{Bmatrix}
 \approx \frac{(-1)^{a+b+c+2(d+e+f+R)}}{\sqrt{2R}}
 \begin{pmatrix}
  a & b & c \\
  e-f & f-d & d-e
 \end{pmatrix}.
\label{asymptotic relation}
\end{align}
Using this formula, one sees that
in the $N_0\rightarrow\infty$ limit 
\begin{align}
\hat{{\cal C}}^{J_1m_1(j'j)}_{J_2m_2(j'j'')\;J_3m_3(j''j)}
\rightarrow {\cal C}^{J_1m_1q_1}_{J_2m_2q_2\;J_3m_3q_3}
\end{align}
with the identification $j'-j=q_1,\;j'-j''=q_2,\;j''-j=q_3$.

The vector spherical harmonics on $S^3$, $S^2$ and the fuzzy sphere are 
defined in terms of the scalar spherical harmonics as 
\begin{align}
&Y_{Jm\tilde{m}i}^{\rho}=
i^{\rho}\sum_{n,p}U_{in}C^{Qm}_{\tilde{Q}p\;1n}Y_{\tilde{Q}p\tilde{m}}, \;\;
\tilde{Y}_{Jmqi}^{\rho}=
i^{\rho}\sum_{n,p}U_{in}C^{Qm}_{\tilde{Q}p\;1n}\tilde{Y}_{\tilde{Q}pq},
\n
&\hat{Y}_{Jm(jj')i}^{\rho}
=i^{\rho}\sum_{n,p}U_{in}C^{Qm}_{\tilde{Q}p\;1n}\hat{Y}_{\tilde{Q}p}^{(jj')},
\label{vector spherical harmonics}
\end{align}
where the unitary matrix $U$ is given by
\begin{align}
U=\frac{1}{\sqrt{2}}
\begin{pmatrix}
-1 & 0 & 1 \\
-i & 0 & -i\\
0  & \sqrt{2} & 0
\end{pmatrix}.
\end{align}
The vector spherical harmonics possess the following properties.

\noindent
\underline{Action of the $SU(2)$ generators}
\begin{align}
&\frac{1}{\mu}\epsilon_{ijk}\nabla_{j}Y_{Jm\tilde{m}k}^{\rho}
=i\epsilon_{ijk}{\cal L}_jY_{Jm\tilde{m}k}^{\rho}+Y_{Jm\tilde{m}i}^{\rho}
=\rho (J+1) Y_{Jm\tilde{m}i}^{\rho}, \nonumber\\
&i\epsilon_{ijk}L_j^{(q)}Y_{Jmqk}^{\rho}+Y_{Jmqi}^{\rho}
=\rho (J+1) Y_{Jmqi}^{\rho}, \nonumber\\
&i\epsilon_{ijk}L_j\circ \hat{Y}_{Jm(jj')k}^{\rho}+\hat{Y}_{Jm(jj')i}^{\rho}
=\rho (J+1) \hat{Y}_{Jm(jj')i}^{\rho}.
\label{action of SU(2) on vector spherical harmonics}
\end{align}
\underline{Complex conjugate}
\begin{align}
&(Y_{Jm\tilde{m}i}^{\rho})^{\dagger}
=(-1)^{m-\tilde{m}+1}Y_{J-m-\tilde{m}i}^{\rho}, \;\;
(Y_{Jmqi}^{\rho})^{\dagger}
=(-1)^{m-q+1}Y_{J-m-qi}^{\rho}, \nonumber\\
&(\hat{Y}_{Jm(jj')i}^{\rho})^{\dagger}
=(-1)^{m-(j-j')+1}\hat{Y}_{J-m(j'j)i}^{\rho}.
\label{complex conjugate}
\end{align}
\underline{Orthonormal relation}
\begin{align}
&\int \frac{d\Omega_3}{2\pi^2}
(Y_{J'm'\tilde{m}'i}^{\rho'})^{\dagger}Y_{Jm\tilde{m}i}^{\rho}
=\delta_{JJ'}\delta_{mm'}\delta_{\tilde{m}\tilde{m}'}\delta_{\rho\rho'},
\nonumber\\
&\int \frac{d\Omega_2}{4\pi}
(\tilde{Y}_{J'm'qi}^{\rho'})^{\dagger}\tilde{Y}_{Jmqi}^{\rho}
=\delta_{JJ'}\delta_{mm'}\delta_{\rho\rho'},
\nonumber\\
&\frac{1}{N_0}\mbox{tr}((\hat{Y}_{J'm'(j'j)i}^{\rho'})^{\dagger}
\hat{Y}_{Jm(jj')i}^{\rho})
=\delta_{JJ'}\delta_{mm'}\delta_{\rho\rho'}.
\label{orthonormal relation for vector spherical harmonics}
\end{align}
\underline{Integral of the product of three harmonics}
\begin{align}
&\int \frac{d\Omega_3}{2\pi^2} \: \epsilon_{ijk}\: 
Y^{\rho_1}_{J_1m_1\tilde{m}_1i} Y^{\rho_2}_{J_2m_2\tilde{m}_2j} 
Y^{\rho_3}_{J_3m_3\tilde{m}_3k}
\equiv {\cal E}_{J_1m_1\tilde{m}_1\rho_1\;J_2m_2\tilde{m}_2\rho_2
\;J_3m_3\tilde{m}_3\rho_3}, \nonumber\\
&\int \frac{d\Omega_2}{4\pi} \: \epsilon_{ijk}\: 
\tilde{Y}^{\rho_1}_{J_1m_1q_1i} 
\tilde{Y}^{\rho_2}_{J_2m_2q_2j} 
\tilde{Y}^{\rho_3}_{J_3m_3q_3k}
={\cal E}_{J_1m_1q_1\rho_1\;J_2m_2q_2\rho_2\;J_3m_3q_3\rho_3}, \nonumber\\
&\epsilon_{ijk}\frac{1}{N_0}\tr
(\hat{Y}^{\rho_1}_{J_1m_1(jj')i} 
\hat{Y}^{\rho_2}_{J_2m_2(j'j'')j} 
\hat{Y}^{\rho_3}_{J_3m_3(j''j)k})
\equiv 
\hat{{\cal E}}_{J_1m_1(jj')\rho_1\;J_2m_2(j'j'')\rho_2\;J_3m_3(j''j)\rho_3}.
\label{integral of product of three vector spherical harmonics}
\end{align}
One can compute ${\cal E}$ and $\hat{{\cal E}}$ using (\ref{C and Chat}).
Their explicit forms are given in \citen{ISTT}. 
In the limit $N_0\rightarrow\infty$, 
$\hat{{\cal E}}_{J_1m_1(jj')\rho_1\;J_2m_2(j'j'')\rho_2\;J_3m_3(j''j)\rho_3}
\rightarrow 
{\cal E}_{J_1m_1q_1\rho_1\;J_2m_2q_2\rho_2
\;J_3m_3q_3\rho_3}$ with the identification
$j-j'=q_1,\;\;j'-j''=q_2,\;\;j''-j=q_3$.

\end{document}